\documentclass[%
 aip,
 amsmath,amssymb,
 reprint,%
]{revtex4-1}
\usepackage{amsmath,amssymb}
\usepackage{dsfont}
\usepackage{graphicx}
\usepackage{hyperref}
\usepackage{physics}
\usepackage{amsmath,amssymb}
\usepackage{dsfont}
\usepackage{graphicx}
\usepackage{hyperref}
\usepackage{physics}
\usepackage{mathtools}
\usepackage{cancel}
\usepackage{bbold}
\usepackage{bm}
\usepackage{color} 
\usepackage[dvipsnames]{xcolor}
\usepackage[T1]{fontenc}
\usepackage{verbatim}

\usepackage{tcolorbox}

\usepackage{soul}

\definecolor{asparagus}{rgb}{0.53, 0.66, 0.42}

\usepackage{ulem} 

\begin{document}
\newcounter{theo}
\author{M. K\k{e}pa}
\affiliation{%
Institute of Physics, Polish Academy of Sciences, al.~Lotnik{\'o}w 32/46, PL 02-668 Warsaw, Poland
}%
\author{{\L}. Cywi\'nski }
\affiliation{%
Institute of Physics, Polish Academy of Sciences, al.~Lotnik{\'o}w 32/46, PL 02-668 Warsaw, Poland
}%
\author{J. A. Krzywda}
 \email{krzywda@ifpan.edu.pl}
\affiliation{%
Institute of Physics, Polish Academy of Sciences, al.~Lotnik{\'o}w 32/46, PL 02-668 Warsaw, Poland
}
\affiliation{%
Lorentz Institute and Leiden Institute of Advanced Computer Science,
Leiden University, P.O. Box 9506, 2300 RA Leiden, The Netherlands
}

\title{Correlations of spin splitting and orbital fluctuations due to 1/f charge noise in the Si/SiGe
Quantum Dot}

\begin{abstract}
Fluctuations of electric fields can change the position of a gate-defined quantum dot in a semiconductor heterostructure. In the presence of magnetic field gradient, these stochastic shifts of electron's wavefunction lead to fluctuations of electron's spin splitting. The resulting spin dephasing due to charge noise limits the coherence times of spin qubits in isotopically purified Si/SiGe quantum dots. We investigate the spin splitting noise caused by such process caused by microscopic motion of charges at the semiconductor-oxide interface. We compare effects of isotropic and planar displacement of the charges, and estimate their densities and typical displacement magnitudes that can reproduce experimentally observed spin splitting noise spectra. We predict that for defect density of $10^{10}$ cm$^{-2}$, visible correlations between noises in spin splitting and in energy of electron's ground state in the quantum dot, are expected.
\end{abstract}
\maketitle

Due to the weakness of spin-orbit coupling in conduction band of silicon, and possibility of isotopic enrichment leading to removal of spinful $^{29}$Si nuclei, a spin of a single electron in an isotopically purified silicon quantum dot (QD) has the longest coherence time of all the QD-based spin qubits \cite{Chan_PRAP18, Yoneda_NN18, Tanttu_PRX19, ZajacScience18, Mills_SA22, philipsUniversalControlSixqubit2022} However, this relative isolation of the spin degree of freedom from external influence makes it hard to perform spin rotations, i.e.~single-qubit quantum gates. The currently commonly accepted solution is to expose such qubits to  magnetic field gradients generated by nearby nanomagnets.\cite{Tokura_PRL06,Neumann_JAP15,Aldeghi_APL23} The gradients of magnetic fields transverse to a global quantization axis, due to constant external $B$ field, allow then for dot-selective electron spin resonance operations driven by ac voltages on gates defining the QDs.\cite{Takeda_SA16, Yoneda_NN18, zhaoSinglespinQubitsIsotopically2019, cornaElectricallyDrivenElectron2018, Struck_NPJQI20}  The gradients of the longitudinal components of the magnetic field lead to qubit-specific spin splittings, further diminishing the addressing errors, when one of two nearby qubits is driven by ac voltages with appropriately chosen frequency.\cite{Li_18SA, Heinz21} However, the presence of the latter gradient makes the splitting of a given spin sensitive to fluctuations of electric fields, as charge-noise-induced fluctuations of an electron position translate into noise in its spin splitting.

Charge noise affecting electron's orbital energy levels has been recently characterized for many silicon-based QDs in heterostructures that are considered as platforms for large-scale semiconductor-based quantum computer.\cite{Connors_PRB19, Connors_NC22, Yoneda22, Rojas23, Struck_NPJQI20} This spurred theoretical investigations of microscopic models of $1/f$ charge noise that could reproduce the observed levels of electron orbital energy fluctuations in SiMOS \cite{Shehata22} and Si/SiGe QDs.\cite{kepa2023} It is commonly assumed that the charge noise in such metal-insulator-semiconductor structures is due to dynamics of charged defects located at the semiconductor/oxide interface.\cite{culcerDephasingSiSinglettriplet2013,Shehata22, Paquelet_Wuetz_2023} Calculations in \onlinecite{Shehata22} and \onlinecite{kepa2023} have established that in order to reproduce the typical order of magnitude of $1/f$ charge noise seen in relevant QDs, the charges should remain trapped at the interface, switching between positions separated by $\delta r \! <\! 1$ nm, i.e.~acting as two-level fluctuators (TLFs). Ranges of densities, $\rho$ of such TLFs, and values of $\delta r$ that lead to reproduction of observed noise, were described in these papers.
Here we follow up on the study from \onlinecite{kepa2023} by adding a magnetic field gradient, and considering fluctuations of spin splitting due to dynamics of sources of charge noise. Comparison of the results of the simulations with measured data on spin dephasing in Si/SiGe QDs exposed to $B$ field gradient, leads to further narrowing down of ranges of parameters of the microscopic models of charge noise that are consistent with state-of-the art experiments.


We use the model of Si/SiGe QD device from \onlinecite{kepa2023}, and start with the model of TLFs described there. Thus we assume typical, constant temperature of $T\approx 100$mK, and equal occupation of TLF states. We consider charges localized at random positions near the semiconductor-oxide interface, with $\rho$ being their interface-averaged planar density, and assume that each of them switches between two positions, $\mathbf{r}_n$ and $\mathbf{r}_n + \delta \mathbf{r}_n$, where the components of $\delta \mathbf{r}_n$ are drawn from independent Gaussian distributions of zero average and  equal variances, such that $\langle |\mathbf{r}_n|\rangle \!= \!\langle\sqrt{\delta x^2 + \delta y^2 + \delta z^2}\rangle \equiv \delta$r is the rms of charge displacement. 



\begin{figure}[tb]
    \centering
\includegraphics[width=\columnwidth]{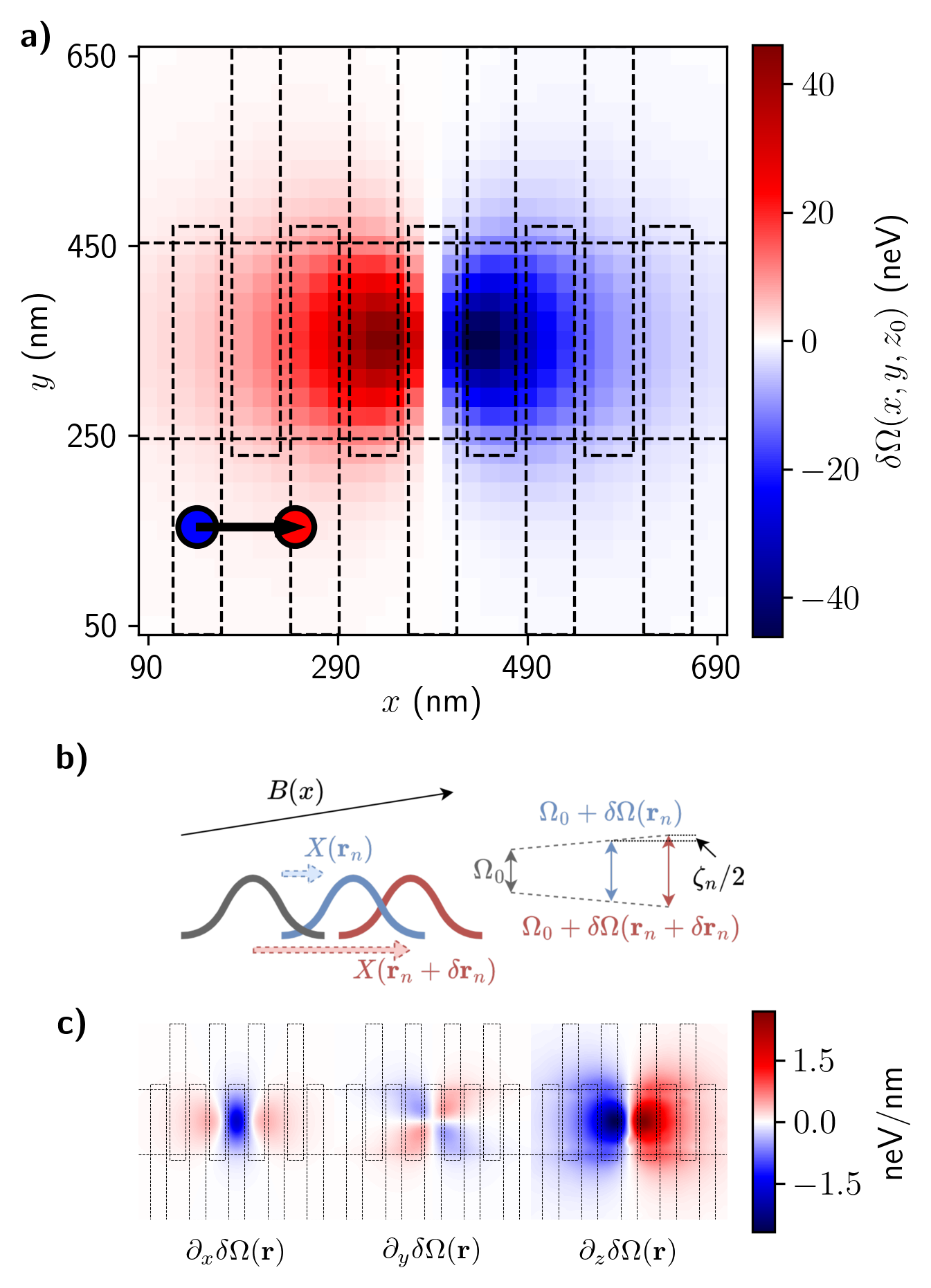}
    \caption{a) Modification of spin splitting $\delta \Omega(x,y,z_0)$ due to presence of a charge defect located close to semiconductor-oxide interface, i.e. at $\mathbf{r} = (x,y,z_0)$. b) Illustration of $\delta \Omega$ caused by the shift of electron wavefunction $\Delta R(\mathbf{r})$ in presence of magnetic field gradient $\Delta B_\parallel = 0.2$mT/nm \cite{}. c) Derivatives along $x$, $y$, $z$ directions used later to compute the TLF-spin coupling $\zeta_n = \nabla \delta \Omega \cdot \delta \mathbf{r}_n$.}
    \label{fig:fig1}
\end{figure}

The model of the device consists of 600x600 nm $\text{Si}_{0.7}\text{Ge}_{0.3}$-Si quantum well (QW) where electrons are trapped, located about 80 nm below the metallic electrodes, and has been implemented in COMSOL (See \onlinecite{kepa2023} for device description). Dashed overlay on Fig.~\ref{fig:fig1}a shows the outline of metallic gates and the channel. We compute the ground and first excited state of the stationary Schroedinger equation in a 10 nm high Si layer of QW, using a built-in COMSOL eigenvalue solver. In this study, the voltage on the plunger gate in the middle of the device is set to 0.25 V, while keeping other gates grounded, giving a realistic orbital gap $\hbar\omega = E_1-E_0 \approx 2$ meV. To study the effect of charged defects, we define a 31x31 regular grid in XY plane. At each point of the grid, and at four different heights $z_0 = 101.75,102.0,102.25,102.5$nm, close to semiconductor-oxide interface, we place a single charge with $q=-|e|$. For each defect position, we extract the expectation values of position operator $\mathbf{R} \equiv \langle \psi |\hat{\mathbf{r}}
|\psi \rangle = \int_V \mathbf{r}|\psi|^2 dv$, where $\psi$ is wavefunction of the ground state and $V$ represents the volume of Si layer. This procedure is followed by the interpolation. 

In presence of longitudinal magnetic field gradient, the displacement of the electron wavefunction is translated to a change in Zeeman splitting,
\begin{equation}
    \delta \Omega(\mathbf{r}_n) = g\mu_{\text{B}} \,\Delta \mathbf{B}_\parallel
\,\cdot \delta \mathbf{R}(\mathbf{r}_n),
\end{equation}
where we used electronic g-factor $g$, Bohr magneton $\mu_B$, $\Delta \mathbf{B}_\parallel$ as gradient of longitudinal component of the magnetic field at the location of the QD, and $\delta \mathbf{R}(\mathbf{r}_n) \! =\! [X(\mathbf{r}_n),Y(\mathbf{r}_n),Z(\mathbf{r}_n)]^{T}$ as the shift of QD position due to presence of the defect at $\mathbf{r}_n$. For simplicity, in this paper we will assume that the longitudinal field  gradient is along the x-axis only, $\Delta\mathbf{B}_\parallel = (\Delta B_\parallel, 0 ,0)^T$, i.e. in the direction parallel to the channel and perpendicular to metallic gates (See Fig. \ref{fig:fig1}). To calibrate our results against experimental results we use here $\Delta B_\parallel = 0.2\,$mT/nm from \cite{Yoneda22, Yoneda_NN18}, which in Si is equivalent to the gradient in Zeeman splitting $\Delta \Omega/\Delta X \approx 0.025 \,\mu$eV/nm. Gradient along a single direction means that the shift in spin splitting is sensitive to the change in $x$-component of QD position only, i.e. $\delta \Omega(\mathbf{r}_n) = g \mu_B \Delta B_\parallel \delta X(\mathbf{r}_n)$.

The raw data are obtained using finite element method implemented in COMSOL software, as described above, and presented in Fig.~\ref{fig:fig1}a for a single plane $z_0 = 102$nm. The resulting shift in Zeeman splitting, $\delta \Omega$, is about four orders of magnitude smaller than the shift in orbital energy, $\delta E$  computed for the same device in \cite{kepa2023}. In contrast to those results, we observe both increase and decrease of $\Omega$, which can be understood as a motion of quantum dot to the right (along the gradient) and to the left (against the gradient) respectively, see Fig.~\ref{fig:fig1}b for the cartoon of the energy shift.

As in \onlinecite{kepa2023}, we employ now the $1/f$ noise model of many independent TLFs corresponding to charges jumping between $\mathbf{r}_n$ and $\mathbf{r}_n +\delta\mathbf{r}_n$, each being a source Random Telegraph Noise characterized by switching rate $\gamma_n$ sampled from a log-normal distribution \cite{schriefl2006decoherence,YouPRR21, Shnirman_PRL05}. We consider switching rates $f_\text{min}<\gamma_n<f_\text{max}$, with experimentally relevant frequencies $f_\text{min} = 10^{-6}$Hz and $f_\text{max} = 10^{6}$Hz. 
We define the coupling between the TLF and the single electron spin qubit as
\begin{equation}
    \zeta_n = \delta \Omega(\mathbf{r}_n+\delta \mathbf{r}_n)- \delta \Omega(\mathbf{r}_n)\approx  \nabla \delta \Omega(\mathbf{r}_n) \cdot \delta \mathbf{r}_n \,\, .
\end{equation}

We assume the contribution from multiple defects is additive, i.e. $\delta \Omega(\{\sum_n \mathbf{r}_n\}) = \sum_n \delta \Omega(\mathbf{r}_n)$. We have tested this assumption against the case of two charges located at random positions, which produced relatively small error $\delta \Omega(\mathbf{r}_1,\mathbf{r_2}) - \delta \Omega(\mathbf{r}_1) - \delta \Omega(\mathbf{r}_2) \ll 0.1\,$neV. This allow us to compute the Power Spectral Density (PSD) of the noise in Zeeman splitting for various charge densities $\rho$, and individual TLFs parameters: displacement size $\delta r_n$, and switching rates $\gamma_n$, without further finite-element simulations. Since each TLF is characterized by a Lorentzian spectrum, the total PSD reads:
\begin{equation}
\label{eq:s_tls}
    S_\text{spin}(f) = \sum_{n=1}^{N} \frac{2\zeta_n^2}{\gamma_n + (2\pi f)^2/\gamma_n} \,\, . 
\end{equation}
where the $n$-the TLF is characterized by  coupling $\zeta_n$ and switching rate $\gamma_n$.
For each realisation, we first use uniform spatial distribution to draw positions of $N$ charges at $z_0 = 102$nm plane, where $N = A\rho$ is related to charge density $\rho$ and considered area $A$. Next, in each realisation, and for each charge we draw the components of displacement vector $\delta \mathbf{r}_n$ from the independent Gaussian distribution of zero average, and the equal variances such that $\langle |\mathbf{r}_n|\rangle = \langle\sqrt{\delta x^2 + \delta y^2 + \delta z^2}\rangle \equiv \delta r$ is the displacement size. 

Using interpolated data from Fig.~\ref{fig:fig1}a, 
we compute the gradients of 
spin splitting shifts
in Fig.~\ref{fig:fig1}c. In all directions we observe that depending on the initial location, the motion of the defect can both increase and decrease $\Omega$. By comparing the relative value of the gradient components, we see the same displacement of a charge along z-direction results in noise larger by about an order of magnitude
in comparison to the in-plane displacements. It is caused by the decrease of the distance between the charge and its image, that is located in the metal. This effectively decreases the dipole moment of charge-image pair. It can be also seen from similarity (with negative sign) between $\delta \Omega$ and $\partial_z \delta \Omega$. In the planar direction, the $x$-derivative is about two times larger than y-, which shows that the motion of the charge along the $x$ direction is more effective at moving the electron's wavefunction along this direction (which is the direction of the gradient of the longitudinal field).

We now translate motion of multiple TLFs into spin splitting noise. We concentrate on slow fluctuations of spin splitting, that directly affect measured coherence time of spin qubit, $T_2^* = \sqrt{2}/\sigma_\text{spin}$, that is parameterized by the noise amplitude
\begin{equation}
\label{eq:sigma}
    \sigma_\text{spin}^2 = 2\int_{f_\text{min}}^{f_\text{max}} S_\text{spin}(f) \text{d}f \equiv \sum_{n} \zeta_n^2.
\end{equation}

\begin{figure}[tb!]
\includegraphics[width=0.9\columnwidth]{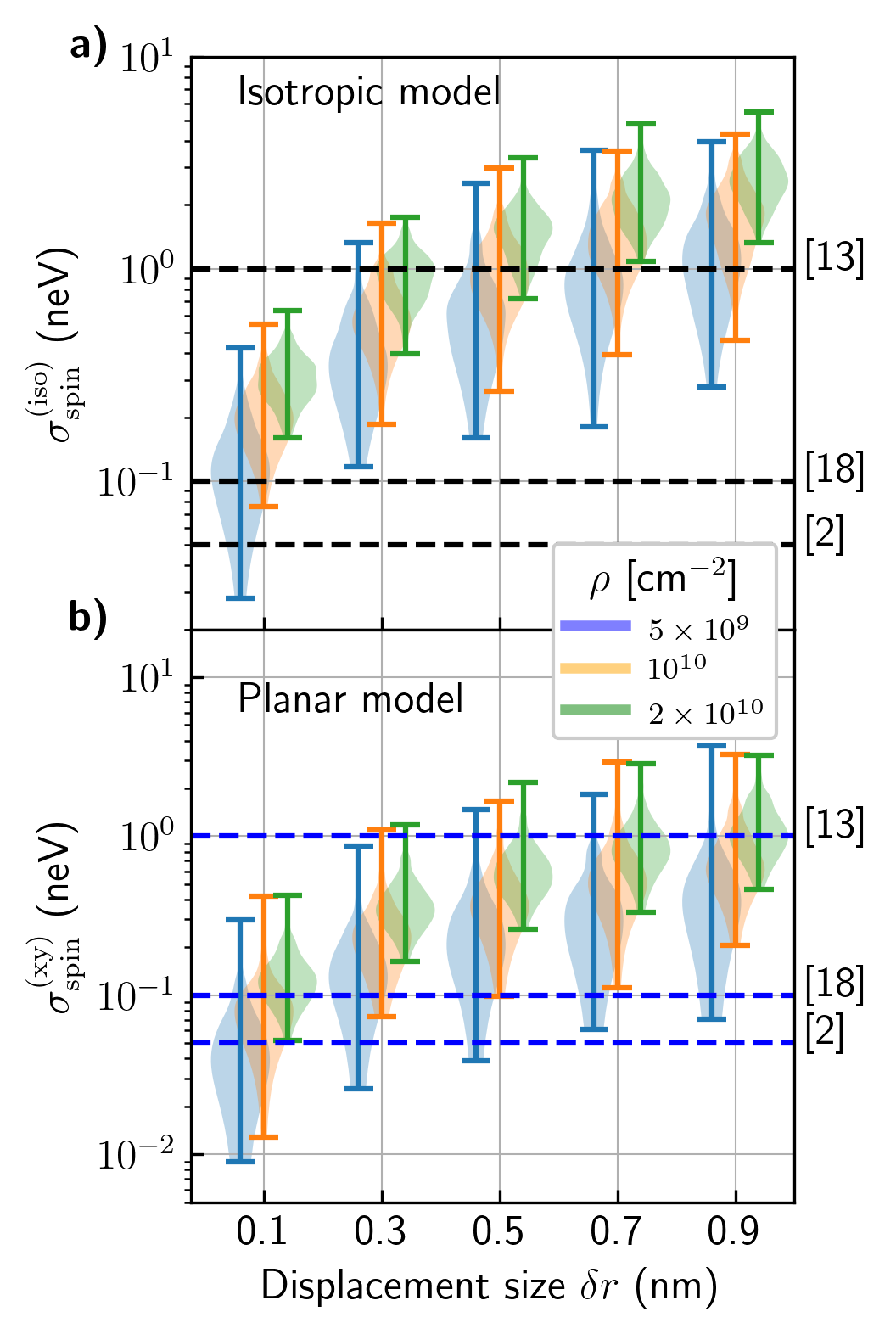}
\caption{Amplitude of spin splitting noise as a function of typical displacement size for three different charge densities in the isotropic model (a) from \cite{kepa2023} and the planar model (b). Violins for given $\delta r$ are slightly offset to increase the legibility of the figure. The numbers on the right correspond to noise amplitude from experimental Ref.~\onlinecite{Yoneda_NN18, Yoneda22,Struck_NPJQI20}.}
    \label{fig:fig2}
\end{figure}

Since typically $T_2^*$ varies between different QDs in isotopically purified Si/SiGe, as a reference we use both the longest-observed coherence time of $T_2^* \! \approx \! 20\mu$s from \cite{Yoneda_NN18,Struck_NPJQI20} and the much shorter $T_2^* \!=\! 1\,\mu$s from \cite{Yoneda22}. Assuming $\Delta B_\parallel \!=\! 0.2\,$mT/nm these values can be related to  $\sigma_\text{spin} = 1\,$neV, (Ref.~\onlinecite{Yoneda22}), $0.05\,$neV (Ref.~\onlinecite{Yoneda_NN18}) and $\sigma_\text{spin} = 0.1\,$neV, (Ref.~\onlinecite{Struck_NPJQI20}), for which we have increased noise amplitude to compensate for twice smaller gradient. In Fig.~\ref{fig:fig2}a we show the statistics of the noise amplitude obtained from 1000 realisations of the isotropic model, for three different charge densities $\rho$ (colors) as a function of displacement size $\delta$r (x-axis). We mark the values of $\sigma_{\text{spin}}$ from references above using dashed lines. For none of the selected pairs of $\rho$ and $\delta r$ it is possible to reconstruct all of these values. Additionally, $\sigma_\text{spin}\!\leqslant\! 0.1\,$neV corresponding to two \onlinecite{Yoneda_NN18,Struck_NPJQI20}, has been reached only for extreme parameters, i.e. the smallest $\delta r=0.1\,$nm, and $\rho = 5\times 10^{9}\,$nm. For larger densities, not more than a single reference value has been reached, independently of $\delta $r.

\begin{figure}[tb!]
    \centering
\includegraphics[width=\columnwidth]{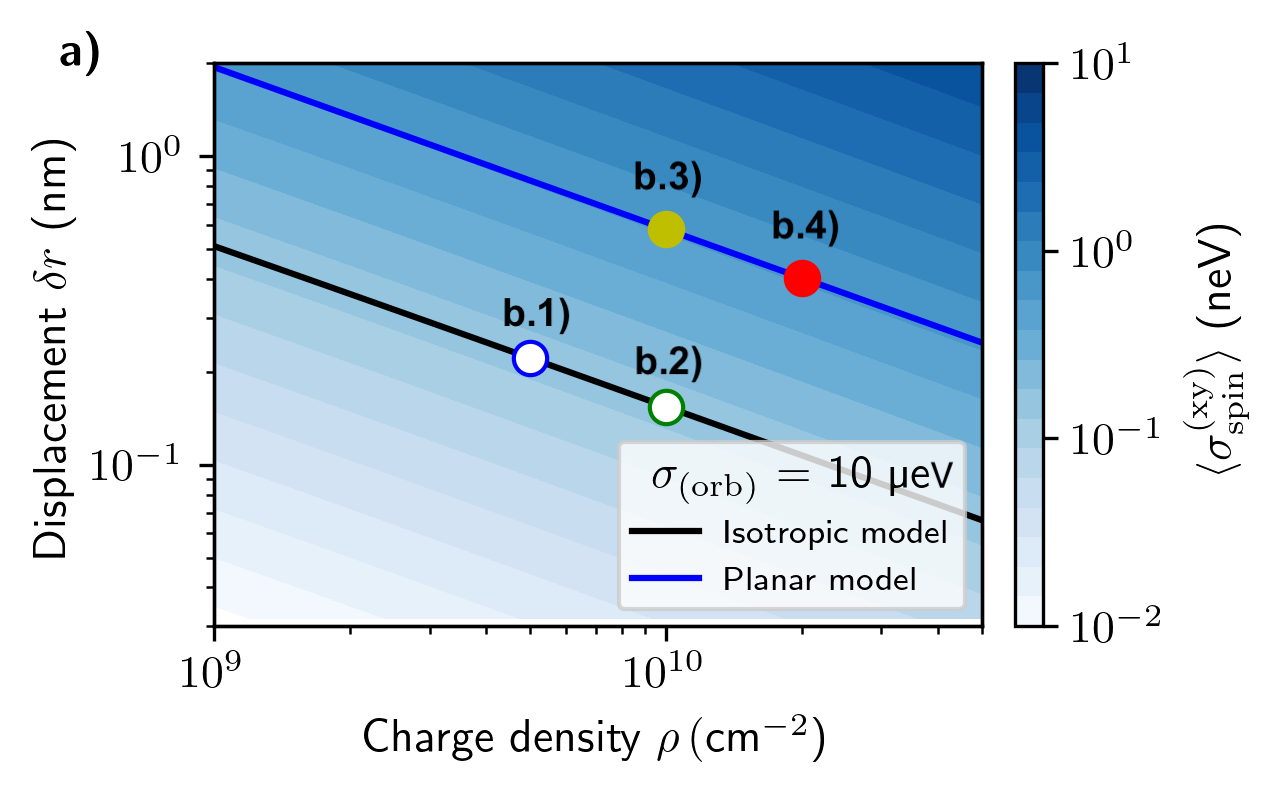}
\includegraphics[width=\columnwidth]{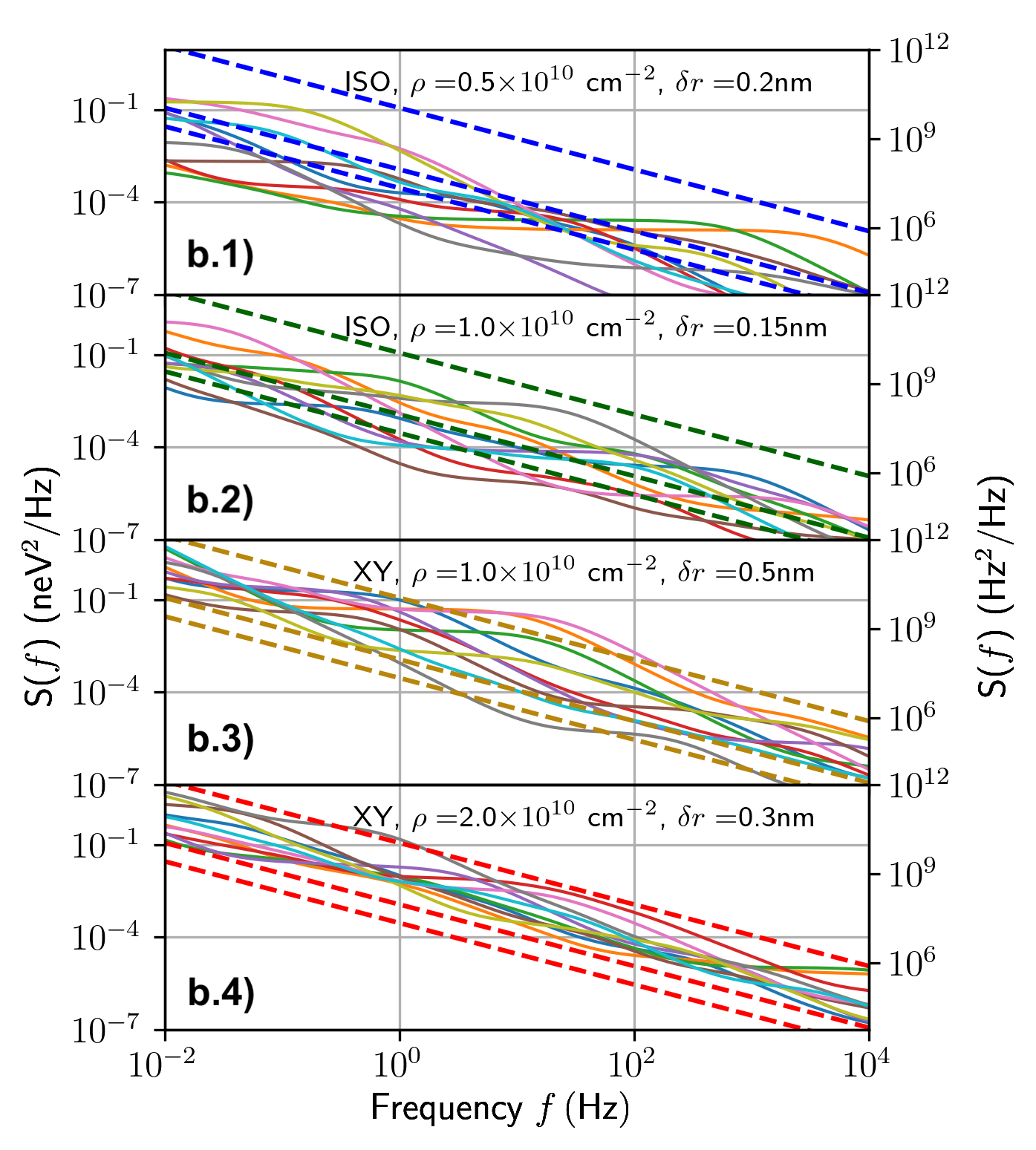}
    \caption{a) Fitted spin splitting noise amplitude $\langle \sigma_{
    \text{spin}}^{(xy)}\rangle$ in the XY model, as a function of charge density $\rho$ and displacement size $\delta $r. Using solid lines we draw a cut through parameter space that corresponds to $\sigma_\text{orb} = 10\mu$eV of orbital energy fluctuations. Filled (hollow) dots correspond to a parameters used in panel b) for planar (isotropic) models. b) 10 realisations of Power Spectral Densities of the spin splitting noise for four parameter pairs corresponding to colored dots in a). For each dot, using the same color, we plot ideal 1/f spectra corresponding to $\sigma_\text{spin} = 0.05,0.1,1$neV using dashed lines.}
    \label{fig:fig3}
\end{figure}

For this reason we consider now an alternative model of charge motion constrained only to the $z_0 = 102$nm plane. We show the result for such a planar motion model in Fig.~\ref{fig:fig2}b. In comparison to the isotropic motion, we observe a small increase of the data spread and, more importantly, a decrease in the average noise amplitude by a factor of about 2. Those changes makes the $\sigma_\text{spin} \! \leqslant\! 0.1\,$neV much easier to achieve for a range of parameters, i.e. $\rho \!\leqslant \!10^{10}\,$cm$^{-2}$ and $\delta r \! \leqslant\! 0.5\,$nm. Simultaneously, some of those parameters allowed also for reaching the largest considered $\sigma_\text{spin} = 1$neV. Thus, a planar model performed better at reconstructing both ends of experimentally relevant values of $\sigma_\text{spin}$. We highlight that relatively larger variance of $\sigma_\text{spin}$, observed in the planar model, can be useful in distinguishing between the two models, given sufficiently large statistics of $T_2^*$ across many experimental devices is collected.

In each model, both ensemble averages and standard deviations of $\sigma_\text{spin}$ follows linear trend with respect to $\delta r$, with their proportionality factor having polynomial dependence on $\rho$ (See Ref.~\onlinecite{kepa2023} for analogous analysis, but for orbital noise). For the average $\sigma_\text{spin}$ we find effective expressions:
\begin{align}
       \langle \sigma_\text{spin}^{\text{(xy)}}\rangle &\approx\bigg[ 0.91 \big(\frac{\rho}{10^{10}\text{cm}^{-2}}\big)^{0.54} + 0.030\bigg]\frac{\delta \text{r}}{\text{nm}} \, \text{neV}. \nonumber \\
       \langle \sigma_\text{spin}^{\text{(iso)}}\rangle&\approx\bigg[ 2.1 \big(\frac{\rho}{10^{10}\text{cm}^{-2}}\big)^{0.53} + 0.025\bigg]\frac{\delta \text{r}}{\text{nm}} \, \text{neV}.
\end{align}
We use the fit from above in Fig.~\ref{fig:fig3}a, where we plot $\langle \sigma_\text{spin}^{(xy)} \rangle$ against $\delta r$ and $\rho$. 

We also perform a cross-check, and compute the amplitude of fluctuation of ground orbital level $\delta E$ within the isotropic and planar model of charge motion. For the isotropic model we use the final result of \onlinecite{kepa2023}, and additionally compute the planar version using $xy$-gradients presented therein. In Fig.~\ref{fig:fig3}a we use the solid lines to plot the cut through parameter space that matches $\langle \sigma_\text{orb} \rangle = 10\mu$eV, corresponding to $T_2^* \! \approx \! 0.1$ns of the charge qubit,\cite{Shi_PRB13} that has been obtained in the isotropic (black) and planar (blue) models of charge motion. On each of them we chose two pairs of parameters $(\rho,\delta r)$. We mark then in Fig.~\ref{fig:fig3}a using colored dots, where hollow (filled) dots corresponds to an isotropic (planar) model. For each of those points, in Fig.~\ref{fig:fig3}b, we plot the corresponding PSD of the spin splitting noise, and compare it with a PSD of ideal 1/f-shape:
\begin{equation}
S_\text{spin}(f) = \frac{\sigma_\text{spin}^2}{2\ln(f_\text{max}/f_\text{min})} \times (1\text{Hz}/f).
\end{equation}
calculated for reference values of $\sigma_\text{spin} = 0.05,0.1,1\,$neV, \cite{Yoneda22,Yoneda_NN18,Struck_NPJQI20}.

In Fig.~\ref{fig:fig3}b we show exemplary 10 realisations of random defect positions $\{\mathbf{r}_n\}$, their (isotropic/in-plane) displacement $\{\delta \mathbf r_n\}$ and switching rates $f_\text{min}<\{\gamma_n\}<f_\text{max}$. We report qualitative agreement between the obtained shapes and experimentally measured spectra,\cite{Yoneda_NN18,Struck_NPJQI20,Yoneda22} which can be seen from visible deviations from 1/f trend due to distinct single TLF features (see also the discussion in \cite{Gungordu_PRB19}). Those features are more prominent in the isotropic model, which can be seen be the direct comparison of the PSDs obtained with $\rho = 10^{10}$cm$^{-2}$ (two middle panels). On the other hand, such a single TLF features are still present after an increase of $\rho$ (last panel). Finally, we observe that quantitatively in isotropic model, parameters $(\rho, \delta r)$ corresponding to $\sigma_\text{orb} = 10\,\mu$eV, gives smaller spin splitting than in the planar model. Also in the latter case, the resulting PSDs are a closer match to the dashed lines, with the prominent example of $\rho = 10^{10}$cm$^{-2}$ and $\delta r=0.5\,$nm, the realisations of which spreads between all reference spectra (dashed lines).

\begin{figure}[tb!]
    \centering
\includegraphics[width=0.9\columnwidth]{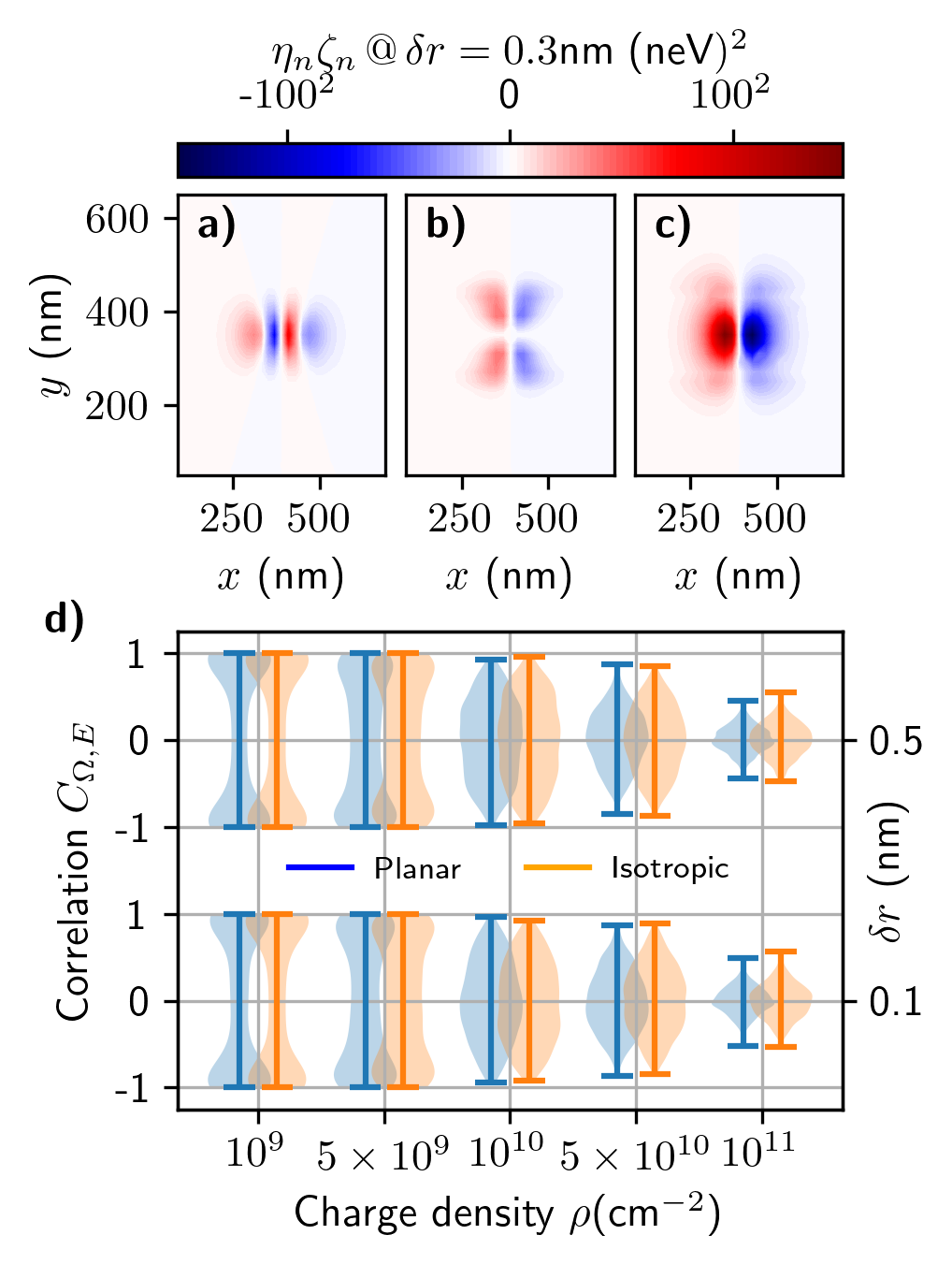}
    \caption{Product of the couplings between a single defect at location $(x,y,z_0=102$nm$)$ and spin ($\zeta_n$) and orbital level ($\eta_n$) for defect displacement size of $\delta r = 0.3$nm along x (a), y (b), z (c) axes. d) Correlation coefficient $C_{\Omega,E}$ between the fluctuations of ground orbital level $E$ and spin splitting $\Omega$ as a function of defect density $\rho$ and for two different displacement sizes $\delta r= 0.1, 0.5$nm. For low-densities two quantities are correlated, and the correlation decreases with the increase of $\rho$.}
    \label{fig:fig4}
\end{figure}

We finally study the correlation between the orbital and spin splitting fluctuations. In analogy to Eq.~\eqref{eq:s_tls} and Eq.~\eqref{eq:sigma} we define the amplitude of low-frequency cross-correlated noise as
\begin{equation}
    \sigma_{\Omega,E}^2 = \sum_n \zeta_n \eta_n,
\end{equation}
 where $\zeta_n$ ($\eta_n$) denotes coupling between the n-th TLF and the spin (ground orbital energy) of the electron. The contribution of each TLF to the correlated noise is then given by the product $\zeta_n \eta_n$. In Fig.~\ref{fig:fig4}a we show such a product for a single TLF located at $\mathbf{r}_n = (x,y,z_0)$ with displacement size $\delta r = 0.3$nm along x-axis (a), y-axis (b), and z-axis (c). We see that for the motion along these three primary directions, both positive and negative correlation of the noise processes are equally probable, which can be seen from the comparison between the sizes of red and blue regions. However, a non-negligible correlation requires the charge to be located in the central region around the QD.

We now analyze relative correlation between orbital and spin splitting noise, defined via 
the normalized covariance:
\begin{equation}
    C_{\Omega,E} = \frac{\sigma^2_{\Omega ,E}}{\sigma_\Omega \sigma_E}.
\end{equation}
where $C_{\Omega,E} = \pm 1$ means perfectly correlated and anti-correlated noises, respectively, and $C_{\Omega,E} = 0$ means that the two fluctuations are uncorrelated.
In Fig.~\ref{fig:fig4}b we plot $C_{\Omega,E}$ as a function of charge density $\rho$ and for two displacement sizes, $\delta r \! =\!  0.1$, $0.5$nm. From their comparison, we conclude that the correlation very weakly depends on $\delta$r. We also confirm that positively- and negatively- correlated noises in $\Omega$ and $E$ are equally probable (note the symmetry with respect to $C_{\Omega,E}=0$). The noise is strongly correlated if the number of TLFs is low, with the extreme case of $N \approx 4$ for $\rho = 10^{9}\,$cm$^{-2}$, and becomes weakly correlated if the number of TLFs is large, with $N = 400$ at $\rho = 10^{11}\,$cm$^{-2}$ as an example. In the former case this is because the noise is typically dominated by a few or even a single TLF located in the central region of Fig~\ref{fig:fig4}a, for which the correlations are large. As $\rho$ increases, both red and blue regions of Fig.~\ref{fig:fig4}a become equally populated, and also contributions from the regions in which the cross-correlations are weak increase. We report that correlations obtained in the isotropic and planar model of charge motion are similar. We also highlight that at the expected densities of $\rho = 10^{10}$cm$^{-2}$, a relatively wide distribution of $C_{\Omega,E}$ is predicted, which means that correlation between $\delta\Omega$ and $\delta E$ is expected to vary visibly between similar devices, or spatially distant regions of the same device. 


In summary, building upon the model of the device and $1/f$ charge noise previously introduced in \onlinecite{kepa2023} we have simulated the influence of charge noise on the spin splitting of an electron in a single quantum dot in a Si/SiGe device, in presence of magnetic field gradient of magnitude used in spin qubit manipulation experiments. 
While in previous modeling\cite{kepa2023} isotropic motion of TLF charges was assumed, here we have introduced a planar model, in which defects were allowed to move only in the plane parallel to the interface. We have found that the planar motion of defects allowed for better agreement with experimentally measured spin splitting noise.  We have predicted values of density of TLFs, $\rho \approx 10^{10}$cm$^{-2}$ and the typical magnitude of motion of charges associated with them, $\delta \! r \! \approx 0.5$ nm, that give results that are consistent with measured orbital energy fluctuations, as well as with other theoretical works.\cite{Rojas23,Shalak22,Shehata22} For this density, noise in spin splitting and in orbital energy should exhibit visible correlations. We believe further distinction between the planar and isotropic model can be made if also spatial correlations of orbital and spin-splitting noise are computed.

\begin{acknowledgments} 
This work was supported by funds from PRELUDIUM grant of the Polish National
Science Centre (NCN), Grant No. 2021/41/N/ST3/02758.
\end{acknowledgments} 

\section*{AUTHOR DECLARATIONS}
\subsection*{Conflict of interest}
The authors have no conflicts to disclose.

\section*{Data Availability Statement}
The data and code that support the findings of this study are available in the dedicated repository \cite{repo}.

\bibliography{literature}
\end{document}